\documentclass[a4paper,11pt]{article}

\usepackage{jcappub} 

\usepackage{qubic}
\usepackage{siunitx}
\usepackage{cancel}
\usepackage{array}
\usepackage{xcolor}
\usepackage[stable]{footmisc}
\usepackage{amsmath}
\usepackage{newtxtext,newtxmath}
\usepackage{placeins}
\usepackage{natbib}
\usepackage{comment}
\usepackage{graphicx}
\usepackage{subfig}
\usepackage{ifthen}
\usepackage{mathtools}

\title{QUBIC VI: cryogenic half wave plate rotator, design and performances}
\author[1,2]{G.~D'Alessandro}
\author[1,2]{L.~Mele}
\author[1,2]{F.~Columbro}
\author[1]{G.~Amico}
\author[1,2]{E.S.~Battistelli}
\author[1,2]{P.~de~Bernardis}
\author[1,2]{A.~Coppolecchia}
\author[1,2]{M.~De~Petris}
\author[3]{L.~Grandsire}
\author[3]{J.-Ch.~Hamilton}
% ...

\affiliation[1]{Universit\a`{a} di Roma - La Sapienza, Roma, Italy}
\affiliation[2]{INFN sezione di Roma, 00185 Roma, Italy}
\affiliation[3]{Universit\'e de Paris, CNRS, Astroparticule et Cosmologie, F-75006 Paris, France}
% ...
\emailAdd{giuseppe.dalessandro@uniroma1.it}
\emailAdd{ lorenzo.mele@roma1.infn.it}
\emailAdd{ fabio.columbro@roma1.infn.it}
\abstract{
{
Inflation Gravity Waves B-Modes polarization detection is the ultimate goal of modern large
angular scale cosmic microwave background (CMB) experiments around the world. A great effort is being spent in the deployment of many ground-based, balloon-borne and satellite experiments using different methods to separate this faint polarized component from the incoming radiation. One of the largely used technique is the Stokes Polarimetry that uses a rotating half-wave plate and a linear polarizer to separate and modulate the polarization components with low residual cross-polarization.
This paper describes the QUBIC Stokes Polarimeter, highlighting its design features and its performances.
The novel design permits to move efficiently the large optical element ($\SI{370}{\milli\meter}$ diameter) at cryogenic temperature, minimizing the large spurious signal due to the emissivity of the optical elements; this constitutes an engineering challenge in order to reduce friction and power dissipation. 
At the designing phase, a particular focus was given to the optimization of the differential thermal contractions between parts.
The rotation is driven by a stepper motor placed outside the cryostat to avoid thermal load dissipation at cryogenic temperature.
The tests and the results presented in this work show that the QUBIC polarimeter can easily achieve a precision below $\ang{0.1}$ in position only using the stepper motor precision and the optical absolute encoder. The rotation induces only few $\SI{}{\milli\kelvin}$ of extra power load on the second cryogenic stage $(\sim \SI{8}{\kelvin})$.

}}
\date{\today}

\begin{document} 

\maketitle


\section{\label{sec:intro}Introduction}
The Q\&U Bolometric Interferometer for Cosmology (QUBIC) is a ground-based experiment which aims to measure with high sensitivity the polarization of the cosmic microwave background (CMB). QUBIC is a bolometric interferometer, a novel concept that combines the calibration control and beam synthesis capabilities of interferometers with the high sensitivity of bolometric detectors. The main goal of the QUBIC instrument is the detection of a characteristic polarization pattern hidden in the CMB, known as B-mode polarization \cite{2020AA.QUBIC.PAPER1}. This predicted, but not yet measured, component is believed to have been imprinted in the CMB by a background of gravitational waves
produced by cosmic inflation via anisotropic Thomson scattering. To date only upper limits have been set on their amplitudes, $r<0.07$ \cite{bicep2018}.
This extremely faint signal, expected to be smaller than a few \SI{}{\nano\kelvin}, is mainly plagued by instrumental systematic effects and by foreground emissions, in particular thermal dust emission from our Galaxy.
QUBIC, as a bolometric inteferometer, can reach the same sensitivity as imagers with the same number of detectors, but it is in addition capable of control and correct systematic effects via the so called \textit{self-calibration} procedure\cite{tartari2016,qubic_2012}.
As every ground-based experiment, QUBIC can perform sky observations only at few frequency bands because of the Earth’s atmosphere, but can mitigate this problem by separating the CMB signal from polarized foregrounds via its spectral-imaging capabilities \cite{2020AA.QUBIC.PAPER2, Mele:proceeding}. Such unique features of the QUBIC instrument allow to extrapolate the signal at frequency sub-bands within the main channels, with a spectral resolution of $\Delta \nu/\nu \sim 0.05$ \cite{2020AA.QUBIC.PAPER2}, helping to separate the cosmological signal from galactic foregrounds.\\
The optical design of QUBIC \cite{2020AA.QUBIC.PAPER8} is optimized around its unique key component, the beam combiner. 

This is the core of the bolometric interferometer and it is composed by the combination of 400 back-to-back corrugated feedhorns, each of them coupled with a controllable blocking blade \cite{2020AA.QUBIC.PAPER7}.
The capability to exclude single back-to-back horn pairs leads to the formation of interference synthesized images on the two orthogonal focal planes suited with arrays of bolometric detectors. Each array has 1024 Transition Edge Sensors (TES) with a noise equivalent power (NEP) of \SI{e-17}{\watt\per\sqrt{\hertz}} \cite{2020AA.QUBIC.PAPER4}. Two frequency bands are selected through a dichroic filter, which reflects radiation at \SI{220}{\giga\hertz} and transmits radiation at \SI{150}{\giga\hertz} on the two focal planes.
The switches will be used to exclude equivalent horn couples, i.e. pairs of horns at the same distance and with the same orientation in the array, which should produce equivalent interference fringes on the focal planes in absence of instrumental systematics. This is the basis of the self-calibration methodology, used to calibrate the instrument with increasing accurancy as more equivalent baselines are detected \cite{2013A&A...550A..59B}.
This procedure provides an extremely powerful tool to disentangle the systematic effects.\\
The extraction with high sensitivity of the polarization angle of the incoming radiation is one of the biggest challenges for CMB polarization experiments. Many strategies have been developed for this purpose, as for the BICEP experiment \cite{BICEP2_2014}, which uses polarization sensitive detectors together with the bore-sight rotation of the whole experiment to retrieve the polarization state of the CMB.
New strategies have been developed, as for the PIXIE experiment \cite{2009AIPC.1141...10B} which mixes the polarization state with the spectral features of the CMB.\\
In QUBIC this is done using a Stokes Polarimeter. This is composed of a retarder plate, commonly an half-wave plate (HWP), which induces a phase shift between the two orthogonal polarization components, coupled with linear polarizer as a polarization selector.
By rotating the HWP it is possible to modulate the Stokes vector's components. 
This polarization modulation methodology is used by a large number of CMB experiments: LSPE \cite{swipe, Lamagna_SWIPE}, ACT-Pol \cite{ACTPol2015}, EBEX  \cite{ebex2017}, SPIDER \cite{spider_hwp_strategy},etc. \\
In this paper, we report the design and the performances of the cryogenic Stokes polarimeter developed for the QUBIC experiment: Section \ref{sec:design} describes the design which allows the rotation at cryogenic temperatures, Section \ref{sec:read-out} shows the custom absolute encoder used to recover the HWP orientation,
while Section \ref{sec:test} describes all the mechanical tests performed in order to improve the mechanics of the device, before its implementation in the QUBIC cryostat. The first tests in the cryostat are described as well in this section.
The sections \ref{sec:thermic_impact} and \ref{sec:polmod} describe the optical and thermal performances of the system.\\
The current version of the instrument is a Technological Demonstrator (TD) \cite{2020AA.QUBIC.PAPER3} composed of a reduced array of 64 back-to-back feed-horns, one-quarter of the \SI{150}{\giga\hertz} detector array, and an HWP with a reduced diameter with respect to the one that will be used in the full instrument.

\section{\label{sec:design}Design}
Nowadays the HWP is one of the most important components in the optical chain of all the instruments which aim at measure the sky polarization with a polarization modulation technique. The equation of a Stokes polarimeter (Eq.~\ref{eq:ideal_polarimeter}) with a rotating HWP allows to separate the first three components of the Stokes vector T, Q, and U:   

\begin{equation}
I = \frac{1}{2}\left[T+Q\cos(4\theta)+U\sin(4\theta)\right].
\label{eq:ideal_polarimeter}
\end{equation}

The HWP can rotate continuously or step-by-step, depending on the experiment scanning strategy. By rotating step-by-step, like in QUBIC, the main issue becomes the HWP emission.
Using low emissivity meta-material \cite{Coughlin:article} can mitigate this systematic. A very efficient way to reduce it consists in keep the HWP temperature as low as possible  \cite{Columbro_systematics}.  
The emission of the HWP at cryogenic temperature, i.e. at \SI{10}{\kelvin}, is around two orders of magnitude lower than an HWP placed at room temperature. 
In principle the HWP can be located at colder temperature but its rotation becomes an issue. Performing a rotation of a large optical element at cryogenic temperature is not easy due to the increase of friction between two co-moving parts. Moreover, different materials, typically used in rotating bearings for instance, contract differently during the cool-down.   
Another possible issue, related to the rotation, is the thermal load dissipated by the rotation. 
A possible solution to this is the implementation of a levitating HWP (adopted by experiments as LiteBIRD\cite{LiteBIRD_HWP}, LSPE\cite{Lamagna_SWIPE, SWIPE_HWP}, EBEX\cite{ebex_pol, ebex_hwp}).
In this configuration the HWP is held by a permanent magnet ring which levitates above a superconductor and it can therefore spin continuously with low friction. 
In QUBIC a step-and-integrate scanning strategy has been adopted/selected \cite{2020AA.QUBIC.PAPER1} as in Spider experiment \cite{spider_hwp_strategy}.
For all these reasons, the QUBIC HWP rotator has been designed to satisfy these requirements: 
\begin{itemize}
 \item scanning strategy requirements: speed $>\ang{3}/s$, orientation accuracy $<\ang{0.2}$;
 \item thermal load lower than $\SI{10}{\milli\watt}$ at temperature of operation.
\end{itemize}

\figurename~\ref{sec:design} shows a CAD 3D model ({\it left}) of the cryogenic HWP rotator and a picture ({\it right}) of the system installed in the QUBIC cryostat. It is placed at the top of the \SI{4}{\kelvin} stage, after the IR blocker filters, as described in \cite{2020AA.QUBIC.PAPER5}. The most important components are the transmission system, the pulley system and the HWP support.
The rotation is produced by an external stepper motor, placed on the top of the cryostat vacuum shell. It is transmitted to the HWP, on the top of the \SI{4}{\kelvin} stage, through two magnetic joints and a fiberglass tube shaft. It is combined with a system of pulleys and a stainless steel belt.
All the parts composing the rotator are described in details in the following subsections. 

\begin{figure}[ht]
\centering
\includegraphics[height=50mm]{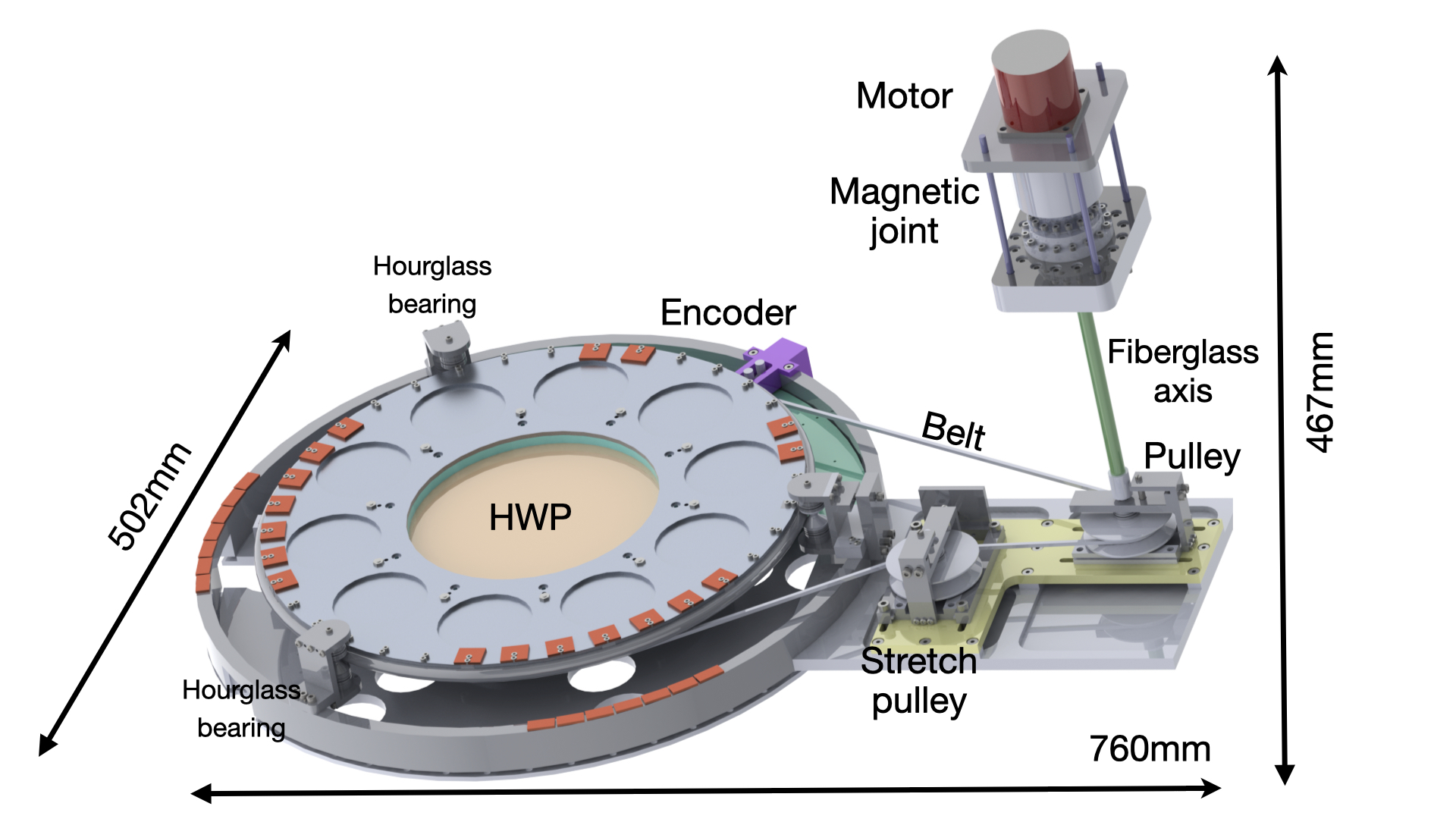}
\includegraphics[height=50mm]{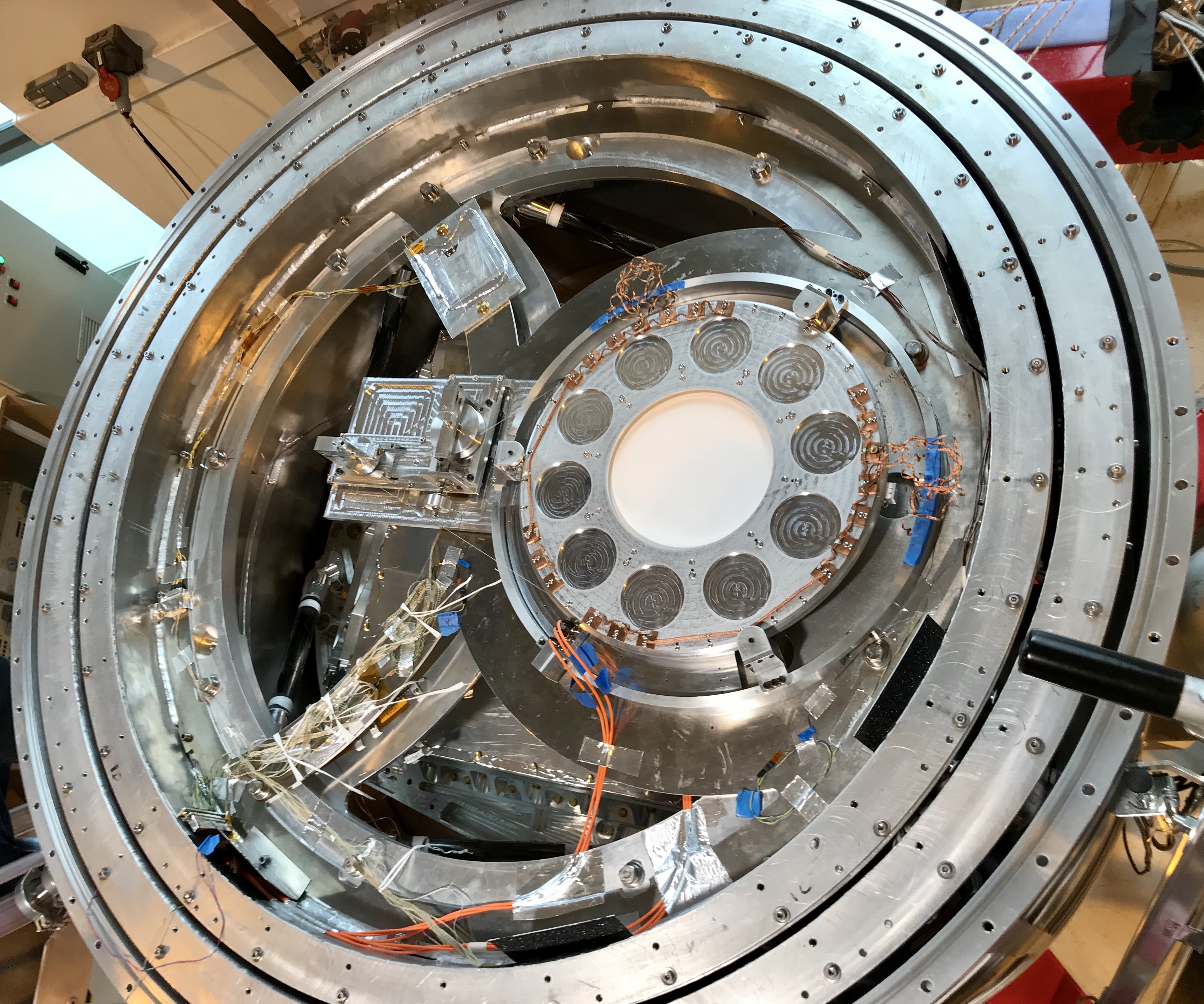}
\caption{\textit{Left}: HWP rotator mechanism renderization. 
\textit{Right}: The HWP rotator mounted inside the QUBIC cryostat for the first tests.}
\label{fig:design}
\end{figure}

\subsection{\label{subse:motor}Motor and transmission} 
In order to reduce the thermal load at the cryogenic stage where the HWP is placed, the motor is placed outside the vacuum vessel at room temperature. The step motor is a Sonceboz 6600-30 Hybrid stepper motor, \SI{1.8}{\degree} per step and $\sim \SI{2}{\newton\meter}$ torque. The transmission ratio, due to the pulleys, between the motor and the HWP support, allows us to move the HWP around of \SI{0.003}{\degree} per step. The motor can be also moved with half-step, in unipolar configuration, and the movement resolution can also be improved.
Since the motor is placed outside the vacuum vessel, we use two magnetic joints which allow to transfer the rotation from the motor to the G-10 fiberglass tube shaft (\SI{140}{\milli\meter} length, \SI{1}{\milli\meter} thick). It transmits the rotation from the motor, at room temperature, to the transmission belt, at \SI{4}{\kelvin} stage. The measured thermal load due to the shaft is a few \SI{}{\milli\kelvin}\footnote{we placed a thermometer directly near the fiberglass axis to evaluate the thermal load. We observed no difference in temperature between this thermometer and the one placed near the HWP support.}.
The transmission belt is a stainless steel braid, $\sim\SI{1.4}{\meter}$ long and $\SI{0.6}{\milli\meter}$ thick. It surrounds the HWP support ring and the pulley system.
In a previous version of the design a Kevlar belt was used, but after several tests in liquid Nitrogen the Kevlar was replaced by the stainless steel belt due to the lower differential thermal contraction with respect to the Aluminum, which is the main material used for the rotation mechanism. 

\subsection{\label{subse:pulley}Pulley system}
The pulley system is composed of two pulleys which help the belt to transmit the rotation from the motor to the HWP support. 
The first pulley is directly connected with the fiberglass shaft. In the first Nitrogen configuration, we noticed that the friction between the belt and this pulley was so little that during the rotation the belt slipped in its groove in the pulley. To improve the friction, the pulley was shaped with a helicoidal groove to make the belt perform at least three revolutions around the pulley.
The second pulley, also called the stretch-pulley, is mounted on sliding support that acts as a tensioner thanks to a spring. The pulley is at equilibrium between the belt and the spring tensions. From the differential thermal contraction point of view, the belt has the largest contribution. This pulley aims to recover all the differential contractions between the belt and the other parts. 
We tuned the spring tension during the test in liquid Nitrogen (\SI{77}{\kelvin}) and the right tension at its nominal temperature inside the cryostat was extrapolated. 
In Section \ref{sec:test} we show how the proper operation of the stretch-pulley was verified while operating inside the cryostat.

\subsection{\label{subse:hwp_supp}HWP support}  
\begin{figure}[ht]
\centering
\includegraphics[height=67mm]{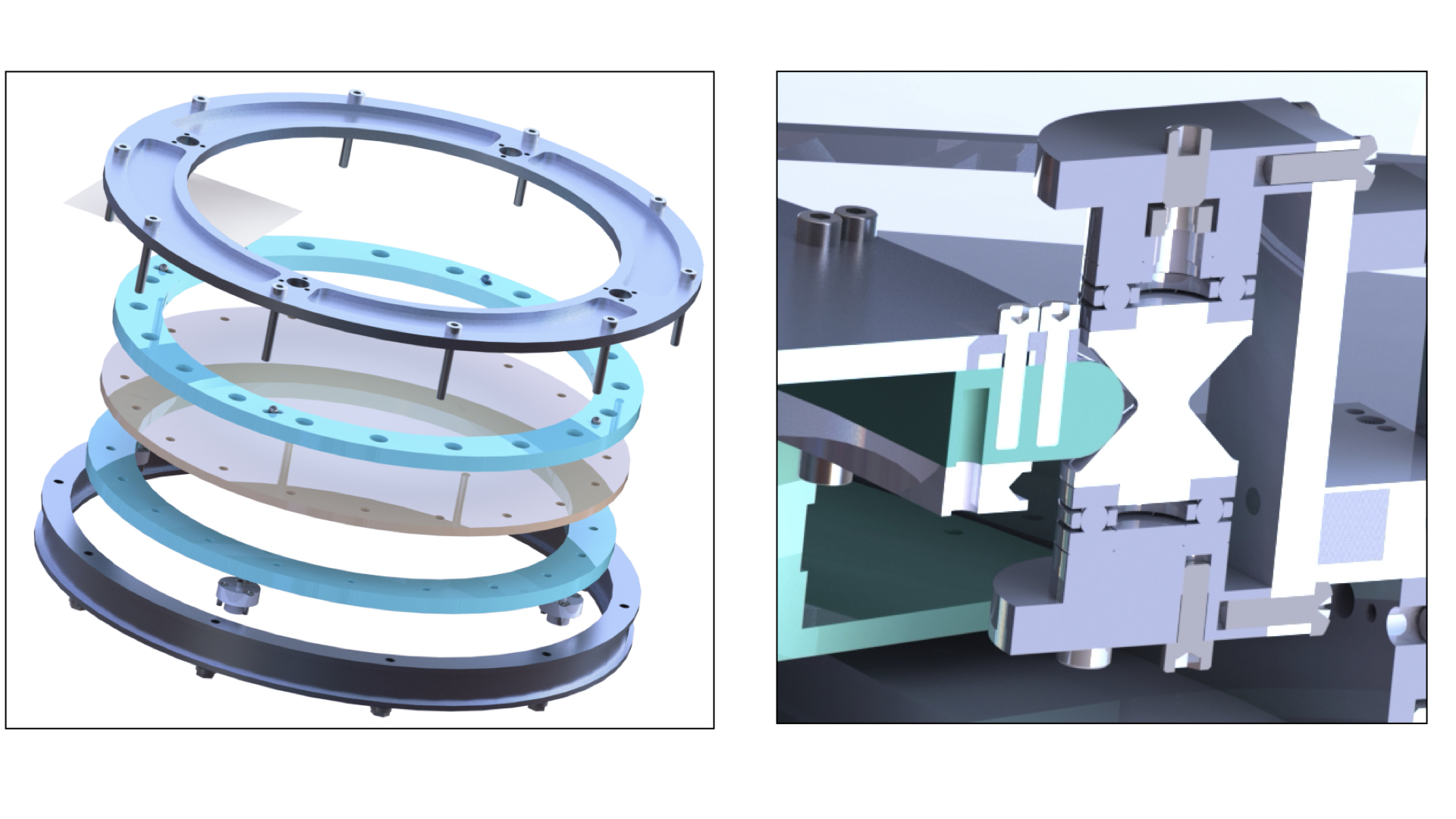}
\caption{\textit{Left}: the HWP hanging system is composed of a PEEK ring, designed in blue, floating in an Aluminum one thanks to 4 spheres. \textit{Right}: hourglass bearing section help to minimize the friction with the HWP support. The thrust bearings are squeezed with elastic washers to leave them free to contract. The HWP support is suspended by three hourglasses equally spaced by \SI{120}{\degree}.}
\label{fig:hang}
\end{figure}

The HWP support is a ring-shaped Aluminum flange suspended by three hourglass-shaped bearings, in \figurename~\ref{fig:hang} (right). The HWP diameter for the Technological Demonstrator instrument is $\SI{180}{\milli\meter}$, while for the Full Instrument it will be $\sim\SI{370}{\milli\meter}$. The Aluminum flange profile has been manufactured with an half- circle shape to produce less friction with the bearings. The hourglasses are squeezed by a couple of thrust bearings fixed with spring belleville washers to leave them free to contract and to prevent them from getting stuck, as shown in the right panel of \figurename~\ref{fig:hang}. We regulate the spring washers tension at \SI{77}{\kelvin} in order to have smooth rotation and reduce the friction inside the bearings.  
An important part of the HWP support is the hanging system, shown in \figurename~\ref{fig:hang} (left): it is not possible to hang the HWP directly by the Aluminum ring flange because they are made of different materials. The HWP is a \SI{4.1}{\milli\meter} polypropylene that has one order of magnitude expansion coefficient greater than the Aluminum. To prevent tensions inside the HWP and changes in its geometry at cryogenic temperature, it has been mounted inside a PEEK\footnote{\url{https://www.victrex.com/~/media/datasheets/victrex_tds_450fe20.pdf}} ring. The PEEK ring is fixed to the Aluminum ring with four stainless steel spheres. The spheres allow a floating contact only in the direction corresponding to an HWP polarization angle equal to zero, preventing orientation angle variations while rotating the device.

\subsubsection{\label{subsubsec:thermalization}HWP thermalization} 
The thermalization of the HWP and its support, described in Section \ref{subse:hwp_supp}, through rigid conductances is not possible, provided that they rotate with respect to the other mechanical parts. 
We used two series of seven elastic copper braids between the HWP support and the nearest fixed ring. The copper braids guarantee a thermal conductivity of \SI{46}{\milli\watt\per\kelvin} at \SI{8}{\kelvin}. Other copper braids surround the HWP support to guaranteed good homogeneity. All the copper braids end with an oxygen-free high-conductivity (OFHC) copper plate fixed with silver glue.  The good thermalization has been tested during the cool down in Nitrogen tank by moving a thermometer from the external side to the internal one.

\section{\label{sec:read-out}Position read-out}
The rotator is a device that can move covering \SI{90}{\degree} of rotation. The readout system is composed of a custom optical encoder, where three pairs of optical fibers face each other. Each couple transmits modulated infrared (IR) light.
The absolute encoder is made by a precisely manufactured metal plate where binary-coded pattern of holes define seven positions. The position of the holes has been made with microns precision and the holes diameter is \SI{2}{\milli\meter}.
The light can be transmitted, when the optical fibers and the holes are aligned, or interrupted when the alignment does not occur. 
The transmitted modulated light travels through the fibers and it is thus detected by photodiodes and processed by the electronics readout. The variable $Bit_N$ is equal to zero or one if the light is interrupted or transmitted. 
The positions are determined by combining the signals from the three photodiodes as in the equation: 
\begin{equation} \label{eq_encoder}
  Pos = \sum_{N=0}^{N=2}  Bit_N \cdot 2^N \ \ \ \ \ Bit_N=0,1  
\end{equation}

This method, already successfully used in PILOT\cite{Salatino_pilot}, allows to separate $\SI{90}{\degree}$ with seven position, $\SI{15}{\degree}$ evenly spaced.  

\subsection{\label{subsec:electronic}Electronic}
The electronics allows to recover the rotator position by reading the signals from the optical fibers of the absolute encoder. The signal is first generated by an oscillator, placed on a Printed Circuit Board, and it is then converted in IR signals by three transmitters, which are directly connected to the three primary optical fibers. The signal can travel from the three primary fibers to the three secondary fibers when the alignment between the fibers and the holes on the gearwheel occurs. The signal from the secondary fibers is, in the end, detected by three photodiodes and lock-in amplified. The combination of the signals from the lock-in amplifiers allows to recover the HWP orientation according to the Eq.~\ref{eq_encoder}.\\
The transmitters\footnote{AVAGO Technologies model HFBR-1412TMZ} wavelenght is optimized at \SI{820}{\nano\meter}. The signal transmitted to the fibers is modulated by the oscillator at \SI{1}{\kilo\hertz}, and it is in the end extracted from uncorrelated noise by a custom synchronous demodulator based on AD630 lock-in amplifiers. The use of a lock-in is fundamental for the system to work properly because of the significant light loss due to the gap of about \SI{7}{\milli\meter} between the primary and the secondary fibers.
The signals from the lock-in amplifiers are processed by a home made firmware running on a Raspberry Pi3 \footnote{\url{https://www.raspberrypi.org}} (RbPi), which recovers the HWP position by continuously reading the lock-in signals. The RbPi also controls the stepper motor, rotating the HWP in accordance with the scan strategy of the sky.
The transmitters and the photodiodes are hosted in an electronics box on a bracket inside the \SI{300}{\kelvin} vacuum shield while the PCB, the motor drive system and the RbPi are placed in an external electronic box.
On the front panel of the external electronic box are placed three LED indicators which show the processed signals from the lock-in amplifiers, thus providing a real time rotator position monitor tool.

\subsection{\label{subsec:software}Firmware}

The firmware was developed in Python\footnote{\url{https://www.python.org}} and runs continuously on the RbPi. 
An User Datagram Protocol (UDP) is used to communicate between Qubic-Studio \cite{2020AA.QUBIC.PAPER3} and the RbPi. The connection is established when the RbPi is turned on and it is controlled by a watchdog every minute. When the connection is established, the RpPi remains on hold until it receives a command from the Qubic-Studio.\
The core of the firmware consists in three main functions which allows to calibrate the rotator in terms of the number of motor steps needed to reach each position and to move the rotator according to the scanning strategy of the sky:

\begin{itemize}
    \item HOME: Find the $1^{st}$ position at $\theta=\SI{0}{\degree}$ from any starting position
    \item CAL \#: Performs \# scans counting the number of motor steps to reach the nominal positions
    \item GOTO \#: Moves the rotator to the position number \# (\# ranges between 1 and 7)
\end{itemize}

The HOME function is necessary since the HWP rotator is fitted on the top of the \SI{4}{\kelvin} stage and it is not visible. The only way to test the rotator operations is through the positions readout system, which provides the HWP orientation when the alignment of the optical fibers with the holes on the gearwheel occurs. The CAL function calibrates the number of motor steps the rotator needs to perform to reach the nominal positions. The calibration function is implemented in the firmware such that the information about the size of single positions can be extrapolated by combining the calibration of the two directions of rotation (a single calibration is meant for both directions of rotation). This is described in details in section \ref{subsubsec:test300_limit}. Given the great reproducibility of the motor steps provided by the calibrations at \SI{300}{\kelvin} and \SI{10}{\kelvin}, as reported in section \ref{subsec:test300} and in section \ref{subsec:test4}, the calibration of the motor steps allows also to move the HWP at intermediate positions, providing a better estimate of the Stokes parameters from the measured polarization modulation curve.\\
After the calibrations of the number of motor steps, the HWP position can be changed with a series of GOTO, which integration time on the single position is imposed by the scanning strategy.
Minor functions are called within the three main functions to set basic parameters of the rotation such as the direction and the velocity, optimized for the scanning strategy.

\section{\label{sec:test}Mechanical test}
The HWP rotator has been tested in several conditions: at room temperature on an optical bench, mainly for the firmware debugging in a safe environment, in liquid Nitrogen, in order to fine tune the strecth-pulley tension at a temperature close to its final working temperature, and , in the end, in its cryostat at a temperature of \SI{10}{\kelvin}. Details of the rotator performance during all these phases are reported in the sections below.

\subsection{\label{subsec:test300}Room temperature}
\begin{figure}
\centering
\includegraphics[width=73mm,height=55mm]{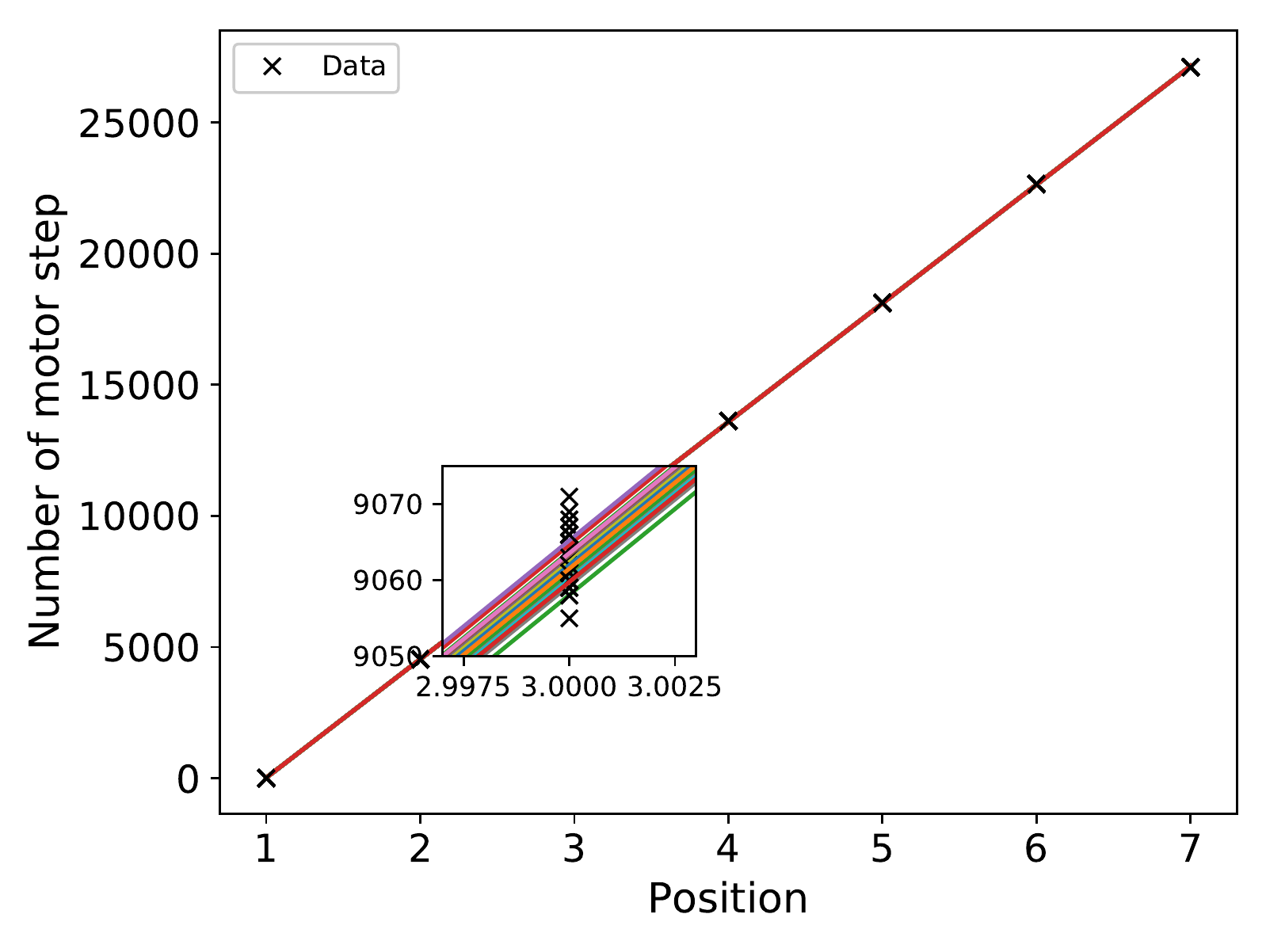}
\includegraphics[width=77mm,height=59mm]{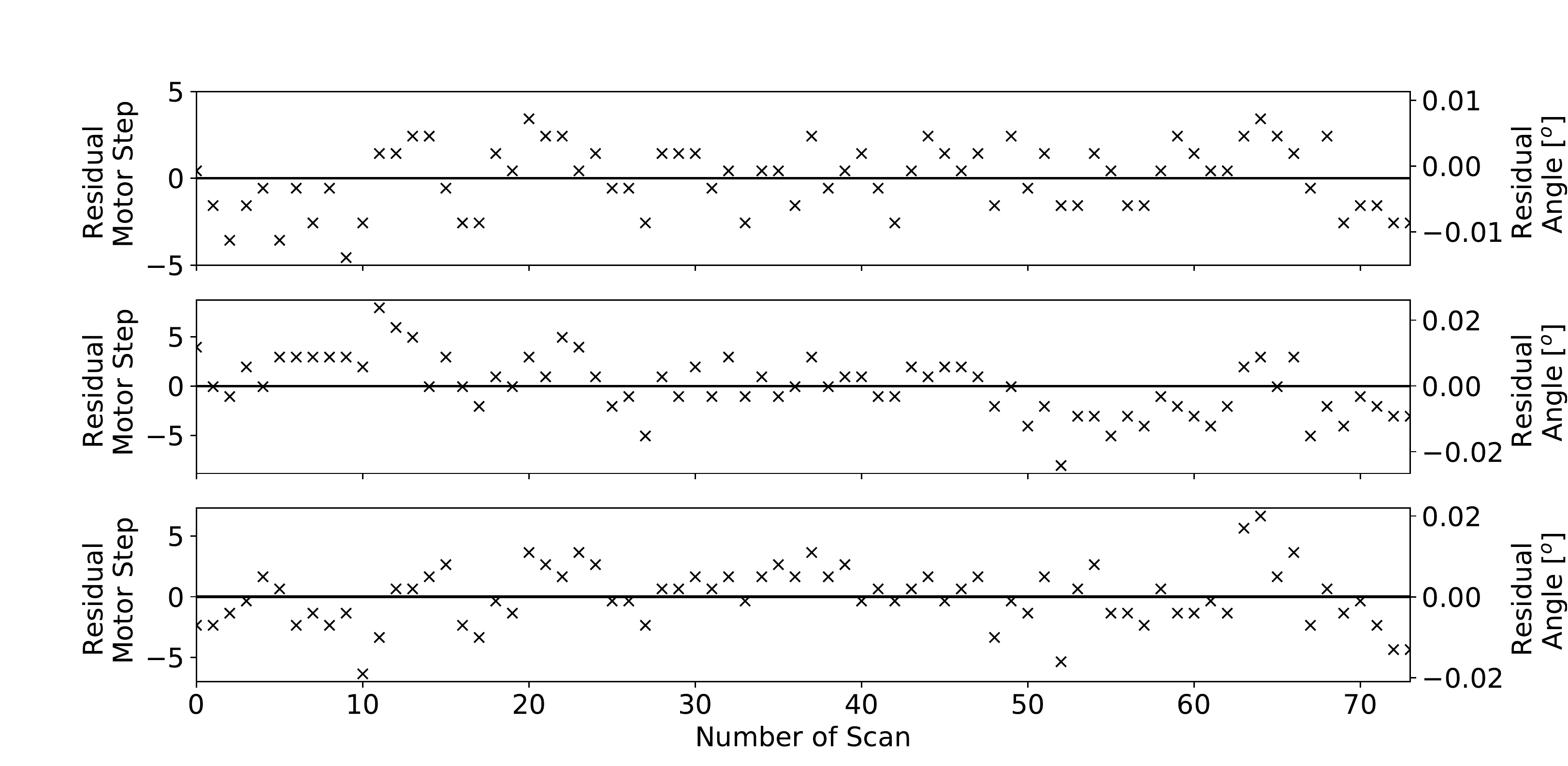}
\includegraphics[width=73mm,height=55mm]{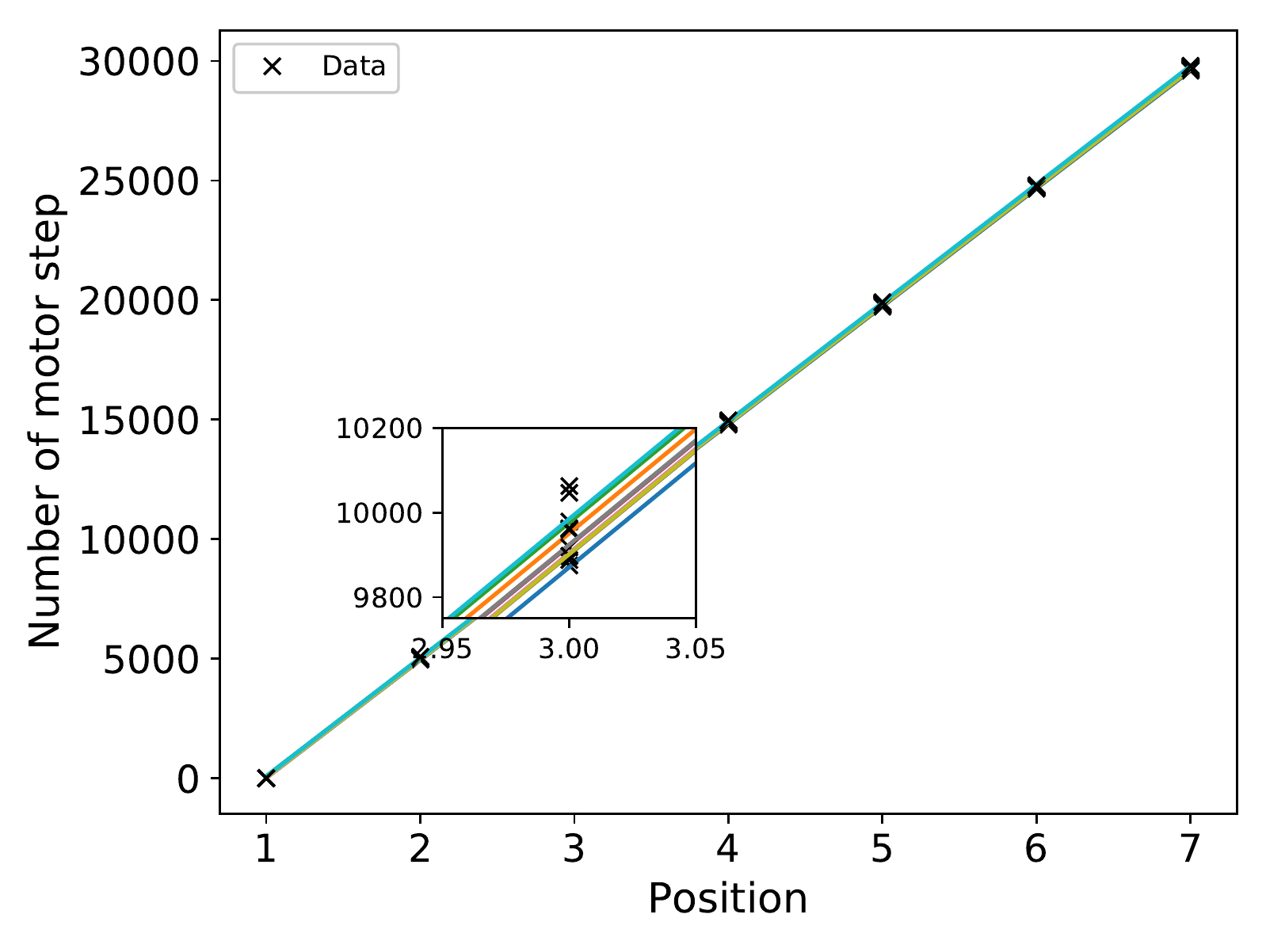}
\includegraphics[width=77mm,height=59mm]{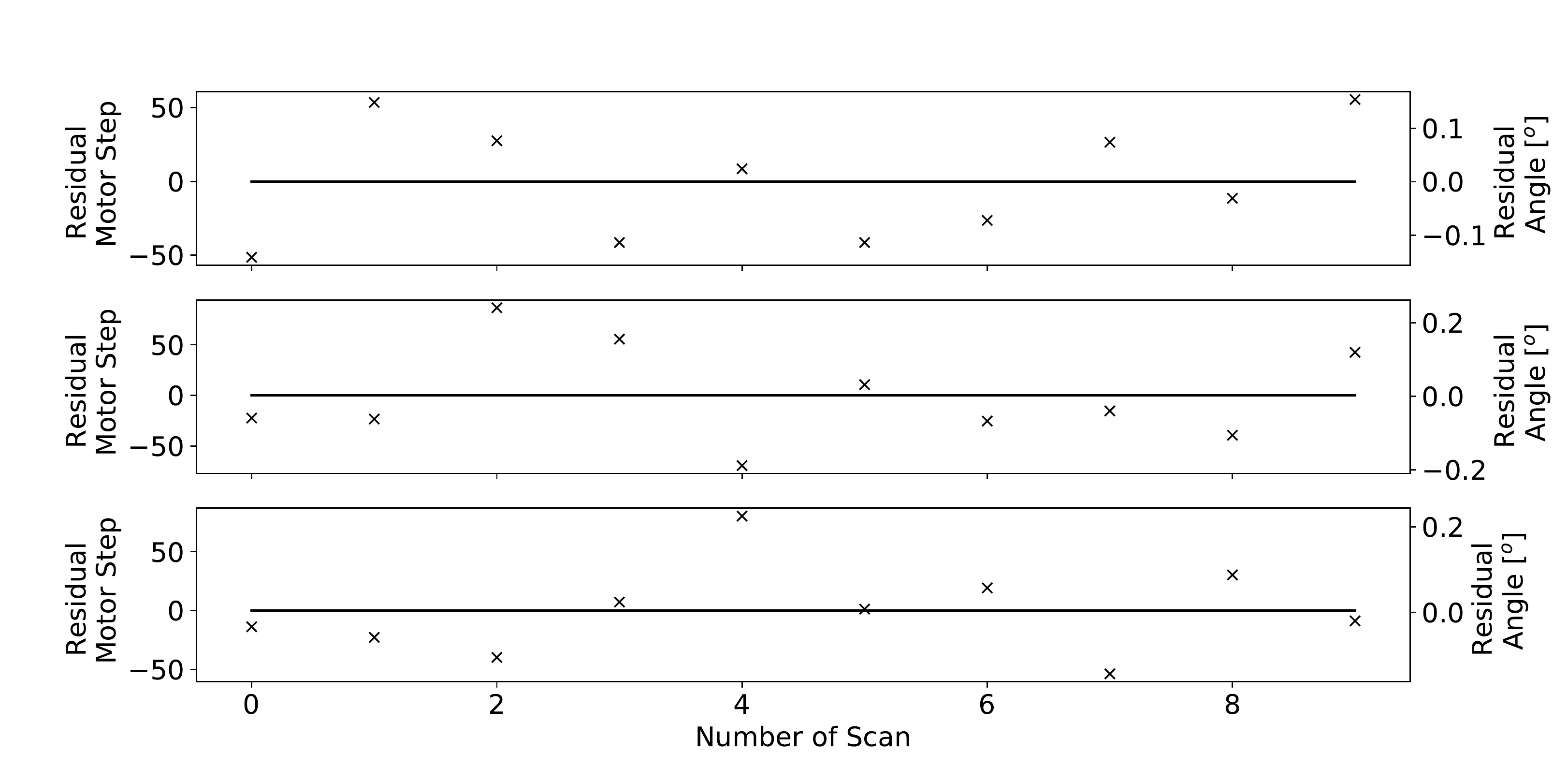}
\caption{
\textit{Left panels}: number of motor steps with respect to rotator positions, with corresponding linear fits, acquired in 75 clockwise scans performed at \SI{300}{\kelvin} (\textit{top}), and acquired during 10 clockwise calibrations performed at \SI{10}{\kelvin} inside the QUBIC cryostat (\textit{bottom}). \textit{Right panels}: residual between the data and their average at each HWP position, reported in terms of motor steps and rotation angles for the 2nd, 3rd and 4th positions as representative cases (from top to bottom sub-panels), for the measurements at \SI{300}{\kelvin} (\textit{top}) and at \SI{10}{\kelvin} (\textit{bottom})
}
\label{fig:calib}
\end{figure}

The HWP rotator has been successfully tested at room temperature, mounted on an optical bench with several hours of operation.
The number of motor steps and the rotator position are acquired automatically by the RbPi. In the top-left panel of \figurename~\ref{fig:calib} we illustrate 75 scans for the clockwise rotation with the corresponding linear fits of data points. In the same figure we show a zoom of the data around the 3rd position to better highlight the great reproducibility achieved. All the curves in this figure overlap within few motor steps ($\sim 10$), where each step corresponds to a rotation of $\SI{0.003}{\degree}$.
The same result is better shown in the top-right panel of \figurename~\ref{fig:calib} as residuals between the motor steps needed to reach the positions 2, 3 and 4 (from top to bottom sub-panels) and their mean value. The residuals are all within $\sim 10$ motor steps. We reported here some representative cases but the same results are achieved for all positions and for the counterclockwise rotation. Given the great reproducibility achieved, it is possible to extract out of these data the information about the size of the rotator positions in terms of motor steps. The details are reported in the following subsection.
The HWP orientation with respect to the optical axis has been measured during the rotation with a Mitutoyo micrometer position sensor, in order to avoid mechanical wobbling of the system and the systematic effect that goes with it \cite{Dale_HWP_wobble}. The planarity of the HWP support has been measured as well to separate the two effects. The wobbling angle in the system is less than $\SI{0.1}{\degree}$.

\subsubsection{\label{subsubsec:test300_limit} Orientation Accuracy at Room Temperature}
\begin{figure}
\centering
\includegraphics[scale=0.5]{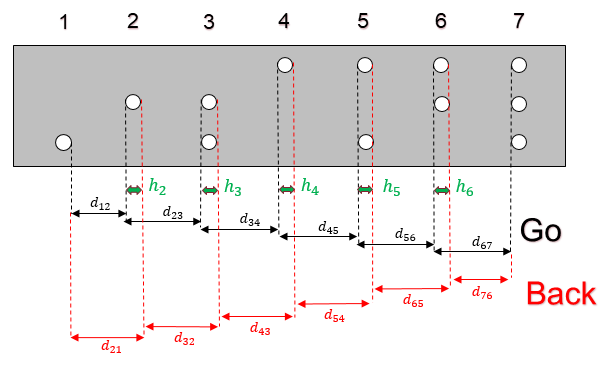}
  \caption{Scheme of the gearwheel equipped with 7 series of holes, defining the rotator positions.  Given the motor steps counts $d_{xy}$ from position $x$ to position $y$, combined from both directions of rotations, the size of intermediate positions can be extracted as the difference between the two scans, highlighted with green arrows in the figure.} \label{fig:4.2}
\end{figure}

Considering the two directions of rotation, we can calibrate our instrument including information about the size of single positions defined by the size of the holes on the gearwheel. Since the count of motor steps is acquired by the RbPi when a position is detected by the readout system, the acquisition occurs at the right edge of the position for the counterclockwise rotation and its left edge for the clockwise rotation. This is illustrated in \figurename~\ref{fig:4.2} as a diagram of the gearwheel with motor steps counts $d_{xy}$ from position $x$ to position $y$.
Combining counts of motor steps of both directions of rotation, we can recover the number of motor steps which are necessary to cross the single position, labeled in \figurename~\ref{fig:4.2} as $h_i$.\\
Since the rotator can be calibrated only between the nominal positions, from the first one at $\theta=\SI{0}{\degree}$ to the last one at $\theta=\SI{90}{\degree}$, this analysis excludes information about the size of the most extreme position, limiting the analysis for the positions ranging between 2 and 6. In Tab.~\ref{tab:4.2} we report the mean amplitude of the $h_i$ terms as degrees of rotation and motor steps with the corresponding error on the mean from 75 calibrations data.
Assuming to rotate the HWP without knowing these measured values of the size of the positions between the 2nd and the 6th, the error one makes in the HWP orientation is constrained by the size of the holes. Since the size of such positions have been measured to be about \SI{0.2}{\degree} - \SI{0.3}{\degree}, to be conservative we can assume that the error made on the orientation is approximately half of the position size, $\sim \SI{0.1}{\degree}$ which already meets the HWP orientation requirement. Assuming that we will be able to reach at \SI{10}{\kelvin} the same repeatability of the rotator positioning that has been achieved at room temperature, we can include the positions size values in the motor step calibration, being therefore able to precisely move the rotator in the center of each position, with an accuracy which is limited by the repeatability that will be actually achieved in the final configuration and at its operation temperature.
The results illustrated here show moreover that, with a suitable calibration of the motor steps, we are also allowed to move the HWP in between two nominal positions (not defined by the gearwheel) with an accuracy that is limited by the repeatability of the final configuration. This will allow a better estimate of the Stokes parameters from a better-sampled polarization modulation curve.

\begin{table}
\centering
\caption{Mean values and errors on the mean for the size of intermediate positions extracted combining data at \SI{300}{\kelvin} from both directions of rotation.}
\begin{tabular}[b]{ccc}
	\hline\hline
	$h_{i}$ & motor steps & degrees\\
	\hline
	$h_2$   &87.7 $\pm$ 0.5 & 0.291 $\pm$ 0.002\\
	$h_3$   &64.9 $\pm$ 0.5 & 0.216 $\pm$ 0.002\\
	$h_4$ 	&115.5 $\pm$ 0.7 & 0.383 $\pm$ 0.002\\
	$h_5$ 	&66.8 $\pm$ 0.8 & 0.222 $\pm$ 0.003\\
	$h_6$ 	&89.9 $\pm$ 0.6 & 0.299 $\pm$ 0.002\\
	\hline

\end{tabular}
	\label{tab:4.2}
\end{table}

\subsection{\label{subsec:nitrogen}Liquid Nitrogen}
In order to verify the right functionality at low temperature, the rotator was cooled down at the liquid Nitrogen temperature (\SI{77}{\kelvin}). The tests have been performed in a big insulated tank where the nitrogen can be transferred. The upper part of the tanks is made of plexiglass allowing a visual inspection during the operation. 
This test allows to calibrate by hand the spring tension at the temperature of \SI{77}{\kelvin}. This calibration should guarantee proper functioning at $\sim\SI{10}{\kelvin}$ (the final temperature of the rotator in the QUBIC cryostat) since the thermal contraction of the aluminum parts stabilizes below $\sim\SI{100}{\kelvin}$ \cite{Duthil:article}.
These tests were very useful to understand the behaviour of the rotator in the cryogenic environment and allowed us to improve its performance thanks to few upgrades:
\begin{itemize}
    \item the kevlar transmission belt has been replaced with one made of stainless steel,
    \item the shape of the first pulley has been changed in order to improve the friction between the pulley and the belt. 
\end{itemize}
By using different magnetic joint we measured a minimum torque of $\sim \SI{0.1}{\newton\meter}$ for an efficient rotation at \SI{77}{\kelvin}. In a worst case scenario where all the power needed to perform the rotation is dissipated by the friction, the thermal load due to the rotation is $\sim \SI{2.6}{\milli\watt}$. 

\subsection{\label{subsec:test4}In QUBIC cryostat}
After these tests, the rotator was installed in its nominal location inside the QUBIC cryostat in order to perform further tests in the right environment. The temperatures detected by two thermometers placed on the HWP support ring and near the HWP reached the equilibrium temperature of about \SI{8}{\kelvin}, \figurename~\ref{fig:thermal}. The tests have been performed by moving successfully the HWP at $\ang{7.5}/s$.\\
\figurename~\ref{fig:calib} (bottom-left) shows ten clockwise rotations acquired during the calibrations performed at cryogenic temperature, together with the corresponding linear fits. The bottom-right panel of \figurename~\ref{fig:calib} shows the residuals at the positions 2, 3 and 4. The measured fluctuations are $\lesssim 100$ motor steps ($\lesssim \SI{0.2}{\degree}$) which are higher than the ones measured at room temperature, but still comparable to the size of the positions measured at \SI{300}{\kelvin}, which determines the orientation error of the HWP at the nominal positions.
We found $\sim 100$ motor steps of backlash at the positions 1 and 7, produced by the steel rope that was not tightened enough to perfectly transmit the rotation. This implies that the stepper motor has to restore the tension of the rope around the primary ring with the first $\sim 100$ steps before starting to transmit efficiently the rotation.
Despite these effects, the performance of the rotator is extremely good, having practically met the requirement for the HWP orientation, provided by the positions size and by the repeatability achieved at cryogenic temperature.

\section{\label{sec:thermic_impact}Thermal performance}
The HWP rotation is transmitted to the HWP support ring thanks to the friction generated in the transmission chain. The motion transmission from the motor ($\SI{300}{\kelvin}$) to the HWP ($\sim\SI{8}{\kelvin}$) is not a totally efficient process because of the friction produced inside the bearings, which is dissipated in the form of thermal power. The effect of the rotation produces a little heating of the cryogenic stage. While the HWP heating is one of the worst systematic effects that has to be mitigated, since it increases the thermal background on the detectors.  
In this section the thermal effect of the rotation is analyzed by means of data acquired by a thermometer mounted nearby the HWP. 
Since a direct measurement of the HWP temperature is not easy to achieve \cite{HWP_temperature} a finite element simulation provides a temperature gradient inside the HWP of about \SI{10}{\milli\kelvin}.    

\subsection{\label{subsec:}Characteristic time constant and heating during the operations}
\begin{figure}
\centering
\includegraphics[scale=0.45]{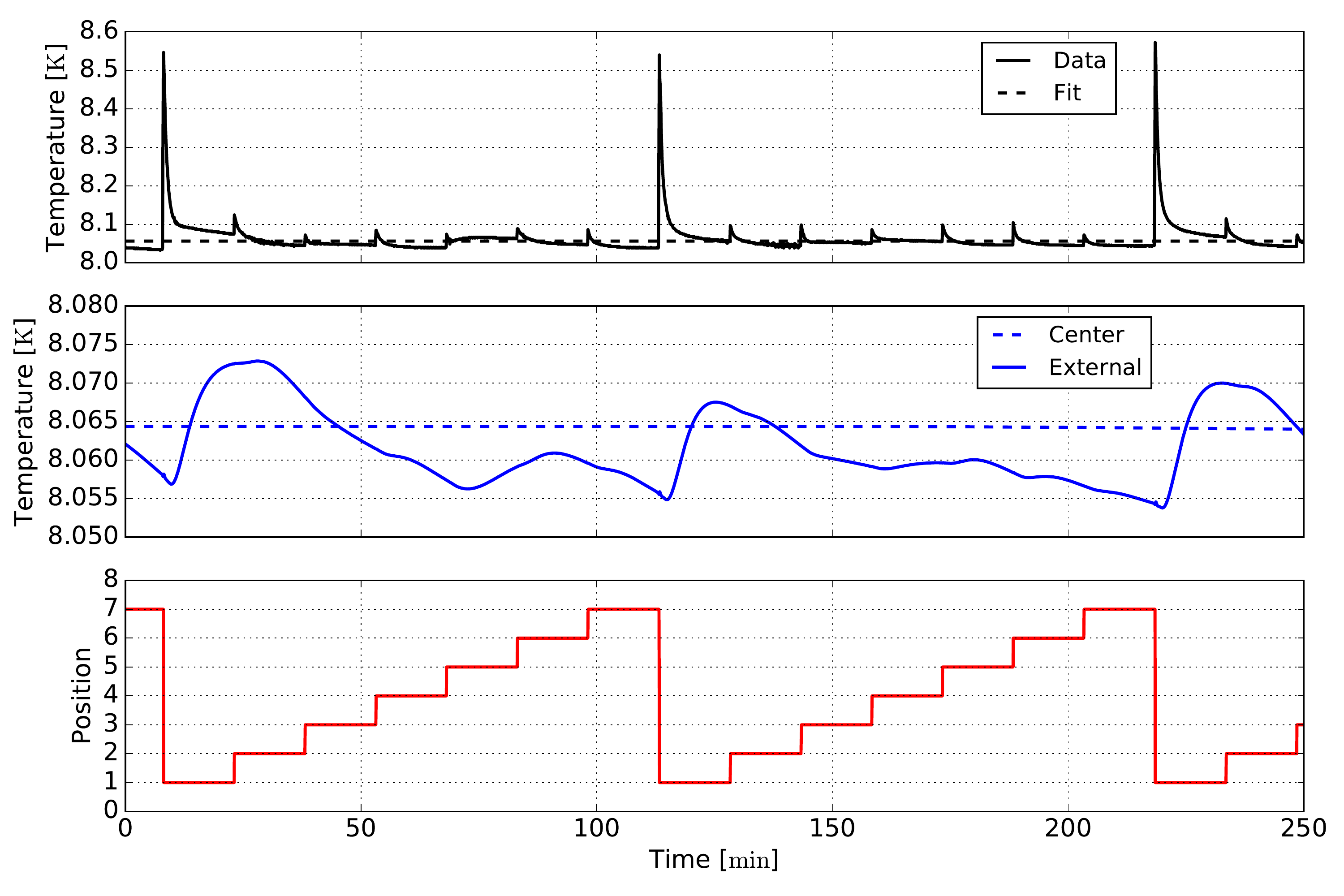}
\caption{\textit{Top panel}: Temperature of the aluminum bearing of the HWP (continuous line) and linear fit (dashed line) of the temperature over a long acquisition time ($\sim\SI{15}{\hour}$) with a slope of $-\SI{1.03}{\milli\kelvin\per\hour}\pm\SI{0.03}{\milli\kelvin\per\hour}$. 
\textit{Central panel}: Finite element simulation for the HWP temperature. In continuum line the external side of the HWP, directly in contact with the PEEK ring; in dashed line the central part of the HWP. We assume the upper part of the PEEK ring has the same temperature of the aluminum bearing showed with continuum line in \textit{Top panel}. 
\textit{Bottom panel}: HWP position read by the absolute encoder.}
\label{fig:thermal}
\end{figure}

Due to the heat produced by the friction during the rotation, the rotator will heat up from their equilibrium temperature, producing a temperature variation with the flowing of time $T(t)$. In order to understand the temperature behaviour of the rotating components, the rotator is moved position by position from the 1st to the 7th, and back directly to position 1 then. Each rotation is performed every $\sim\SI{15}{\minute}$ while the whole test lasts more than half a day. A thermometer (Lakeshore DT-670) was connected to the the aluminun bearing of the HWP. During these scans the thermomether mounted on the \SI{4}{\kelvin} stage, nearby the rotator, does not measure any significant variation. The temperature $T_B \sim \SI{8}{\kelvin}$ of the stage is therefore assumed as a constant. The \textit{top panel} of \figurename~\ref{fig:thermal} shows the temperature detected nearby the HWP (continuous black line) taken during 2 sample scans while the \textit{bottom panel} shows the position data read by the absolute encoder. The change of the position is related to a jump in the temperature profile, which reaches quickly a temperature $T^*$ and, after the position is reached, starts to drop down again with an exponential decay.
We analyzed a long term variation of the rotator temperature with a linear fit performed over a long acquisition time ($\sim\SI{15}{\hour}$). This linear fit (black dashed line in \textit{Top panel} of \figurename~\ref{fig:thermal}) has a slope of $-\SI{1.03}{\milli\kelvin\per\hour}\pm\SI{0.03}{\milli\kelvin\per\hour}$ which means that the rotator is not heating up during the scans but it is still slowly cooled down by the pulse tubes.\\
The thermal model which describes the cool-down of the rotator after the heating induced by the motion follows the thermodynamic equation:
\begin{equation}
\label{eq:thermal_decay}
C \frac{d(T(t) - T_B)}{dt} = - K (T(t) - T_B)
\end{equation}
where $C$ is the heat capacity of the rotator and $K$ is the thermal conductance multiplied by the cross-sectional surface area between the rotator and the \SI{4}{\kelvin} stage.\\
Data were interpolated with the following exponential solution:
\begin{equation}
T(t) - T_B = C^* e^{-\frac{Kt}{C}}
\end{equation}
where $C^*$ is related to the temperature variation induced during the heat up of the rotator and $\tau = \frac{C}{K}$ is the characteristic time constant. \\
Since the scanning strategy of QUBIC consists in moving the HWP from a position to the closest one (i.e. position by position from 1 to 7 and then back again to position 1 in the same way), the small temperature jumps are the most representative ones for the thermal behaviour of the rotation system.
A weighted average of the obtained time constants ($\sim 40$ fits for 8 scans) provides the best value of $\tau = \SI{28.6}{s} \pm \SI{0.6}{\second}$.\\
During the rotation the Eq.~\ref{eq:thermal_decay} becomes:
\begin{equation}
\label{eq:thermal_equation}
C \frac{d(T(t) - T_B)}{dt} = K (T(t) - T_B) + \dot{Q}_M
\end{equation}
where $\dot{Q}_M$ is the heating power induced by the rotation. In principle this value depends on the rotation speed which is kept approximately constant during this test.
The solution of the thermal equation is:
\begin{equation}
(T(t) - T_B) = C^* e^{\frac{Kt}{C}} - \frac{\dot{Q}_M}{K}
\end{equation}

The average heating for a single step of the rotator is $\Delta T = \SI{33}{} \pm \SI{7}{\milli\kelvin}$, a small variation respect to its temperature. If needed this value could be reduced by lowering (via software) the speed of the rotator and so the power dissipated.


\subsubsection{\label{subsec:HWP_temperature}HWP temperature}
In order to determine the temperature of the HWP, we set up a finite element thermal model in COMSOL Multiphysics\footnote{\url{https://www.comsol.com}}. 
The PEEK ring which hosts the HWP is heated up with the measured temperature profile ({\it top panel} of \figurename~\ref{fig:thermal}) and the thermal contact with the HWP is assumed perfect. 
The HWP was assumed to be made by polypropylene and its thermal capacity and specific heat as a function of temperature were obtained by \cite{BARUCCI2002551,RUNYAN2008448}. 
We have also taken into account the radiation heat load produced by the \SI{8.5}{\kelvin} filters which were absorbed by the HWP with an absorption coefficient $\xi = 0.03$.
{\it Middle panel} of \figurename~\ref{fig:thermal} shows the expected temperature of the HWP in its center and at the edge, close to the PEEK ring. The maximum temperature difference between these two points is $\SI{10}{\milli\kelvin}$.


\section{\label{sec:optic}Optical features}
The QUBIC Stokes polarimeter is composed by a polypropylene meta-material HWP followed by a polarizer, where copper is evaporated on a Mylar substrate.
In the QUBIC-TD the HWP and the polarizer have a reduced diameter with respect to the full-instrument configuration, $\SI{180}{\milli\meter}$ and $\SI{370}{\milli\meter}$ respectively.
The HWP has been characterized with dedicated laboratory measurements and the Mueller matrix has been calculated at $\SI{150}{\giga\hertz}$.
In this section we retrieve a model for the polarization modulation and its validity at $\SI{150}{\giga\hertz}$ is discussed.   

\subsection{A model for Stokes polarimeter of QUBIC-TD}
First, a vector network analyzer is used to extract the measured Jones Matrix of the HWP in the range $\SI{110}{\giga\hertz}$-$\SI{170}{\giga\hertz}$. The mean between $\SI{145.5}{\giga\hertz}$ and $\SI{150.5}{\giga\hertz}$ provides: 
$$
J_{HWP}=\left[\begin{matrix}
0.9584 & <0.0001 \\
<0.0001 & -0.9589   
\end{matrix}\right]
$$
The corresponding Mueller matrix can be easily obtained from Eq.~\ref{eq:Muller_to_Jones}: 
\begin{equation}
\label{eq:Muller_to_Jones}
M_{ij}= tr(\sigma_i \cdot J \cdot \sigma_j \cdot J^{\dagger})
\end{equation}
where $\sigma_n$ ($n=[0,...,3]$) are Pauli matrices \cite{Anderson:94}.
The HWP Muller matrix is: 
$$
M_{HWP}=
\left[\begin{matrix}M_{TT} & M_{TQ} & M_{TU} & M_{TV}\\M_{TQ} & M_{QQ} & M_{QU} & M_{QV}\\M_{TU} & M_{QU} & M_{UU} & M_{UV}\\M_{TV} & M_{QV} & M_{UV} & M_{VV}\end{matrix}\right]=
\left[\begin{matrix}0.92 & <10^{-3} & <10^{-7}  & 0\\ <10^{-3}  & 0.92 & <10^{-3}  & 0\\ <10^{-7} & <10^{-3} & -0.92 & 0\\  0 & 0 & 0 & -0.92\end{matrix}\right]
$$
The out-of-diagonal terms, related to temperature to polarization leakage, $M_{IQ}$, $M_{IU}$ and, polarization to polarization leakage, $M_{QU}$, are always less then $10^{-3}$ at $\SI{150}{\giga\hertz}$. We therefore neglect their contribution in the rest of the analysis.\\
The polarizer is a $10 \mu m$ period lithographically etched copper wire device.  The ‘wires’ are  $5\mu m$ wide and 400 nm thick;  the substrate is $1.9 \mu m$ Mylar.
The mylar optical features at $150GHz$ are: 
$n=1.830 + 0.03i$, $\epsilon=3.35$, $ \tan\delta = \SI{e4}{}$ \cite{mylar_feature}. The associated Mueller matrix for zero degree incident angle is: 
$$
M_p{}_x=\frac{1}{2}\left[\begin{matrix}P_{x} & P_{x} & 0 & 0\\P_{x} & P_{x} & 0 & 0\\0 & 0 & 0 & 0\\0 & 0 & 0 & 0\end{matrix}\right]=\frac{1}{2}\left[\begin{matrix}0.9995 & 0.9995 & 0 & 0\\0.9995 & 0.9995 & 0 & 0 \\0 & 0 & 0 & 0\\0 & 0 & 0 & 0 \end{matrix}\right]
$$
We can now provide the Stokes polarimeter model for the QUBIC-TD as follows:
\begin{equation}
    s_{out}= M_{px} \cdot M_{rot}(\theta)^{-1} \cdot M_{HWP} \cdot  M_{rot}(\theta) \cdot s_{in}
\end{equation}
where $s_{in}=(T,Q,U,V)$ is the generic input Stokes vector and $M_{rot}(\theta)$ is a rotation by an angle $\theta$ with respect to the optical axis.
The first component of $s_{out}$ is the intensity at the detector, modulated by the HWP orientation $\theta$: 
\begin{eqnarray}
I_r = \frac{1}{2}T \left(M_{TQ} P_{x} \cos{\left(2 \theta \right)} + M_{TT} P_{x} - M_{TU} P_{x} \sin{\left(2 \theta \right)}\right) + \nonumber \\
\frac{1}{2}Q \big( \left( M_{QQ} P_{x} \cos{\left(2 \theta \right)} - M_{QU} P_{x} \sin{\left(2 \theta \right)} + M_{TQ} P_{x} \right) \cos{\left(2 \theta \right)} - \nonumber \\
\left(M_{QU} P_{x} \cos{\left(2 \theta \right)} + M_{TU} P_{x} - M_{UU} P_{x} \sin{\left(2 \theta \right)}\right) \sin{\left(2 \theta \right) } \big) + \nonumber  \\ 
\frac{1}{2}U \big( \left(M_{QQ} P_{x} \cos{\left(2 \theta \right)} - M_{QU} P_{x} \sin{\left(2 \theta \right)} + M_{TQ} P_{x}\right) \sin{\left(2 \theta \right)} \nonumber \\  
+ \left(M_{QU} P_{x} \cos{\left(2 \theta \right)} + M_{TU} P_{x} - M_{UU} P_{x} \sin{\left(2 \theta \right)}\right) \cos{\left(2 \theta \right)} \big) + \nonumber \\  
\frac{1}{2}V \left(M_{QV} P_{x} \cos{\left(2 \theta \right)} + M_{TV} P_{x} - M_{UV} P_{x} \sin{\left(2 \theta \right)}\right)
\end{eqnarray}
Since $M_{TV}=M_{QV}=M_{UV}=0$, and neglecting the out-of-diagonal terms $M_{TQ}$, $M_{TU}$, $M_{TV}$, $M_{QU}$, $M_{QV}$ and $M_{UV}$ since they are always $<10^{-3}$, the intensity at the detector becomes:
\begin{equation}
I_r = \emph{eff} \cdot  \frac{1}{2}\left[T+Q\cos(4\theta)+U\sin(4\theta)\right] 
\label{eq:qubic_150_modul}
\end{equation}
where the efficiency factor is $\emph{eff}\sim0.92$. All the other terms in the model are less than $10^{-3}$, we can therefore neglect these components in this paper, being at much lower orders of magnitude with respect to the main components. 
We will use all the information related to the Stokes polarimeter Mueller matrices as a function of frequency, where it is not always possible to neglect the out-of-diagonal terms, in a preparation paper about polarization modulation forecast for QUBIC \cite{Dale_qubic_polari}.

\section{\label{sec:polmod}Polarization modulation}
\begin{figure}[ht]
\centering
\includegraphics[width=80mm,height=60mm]{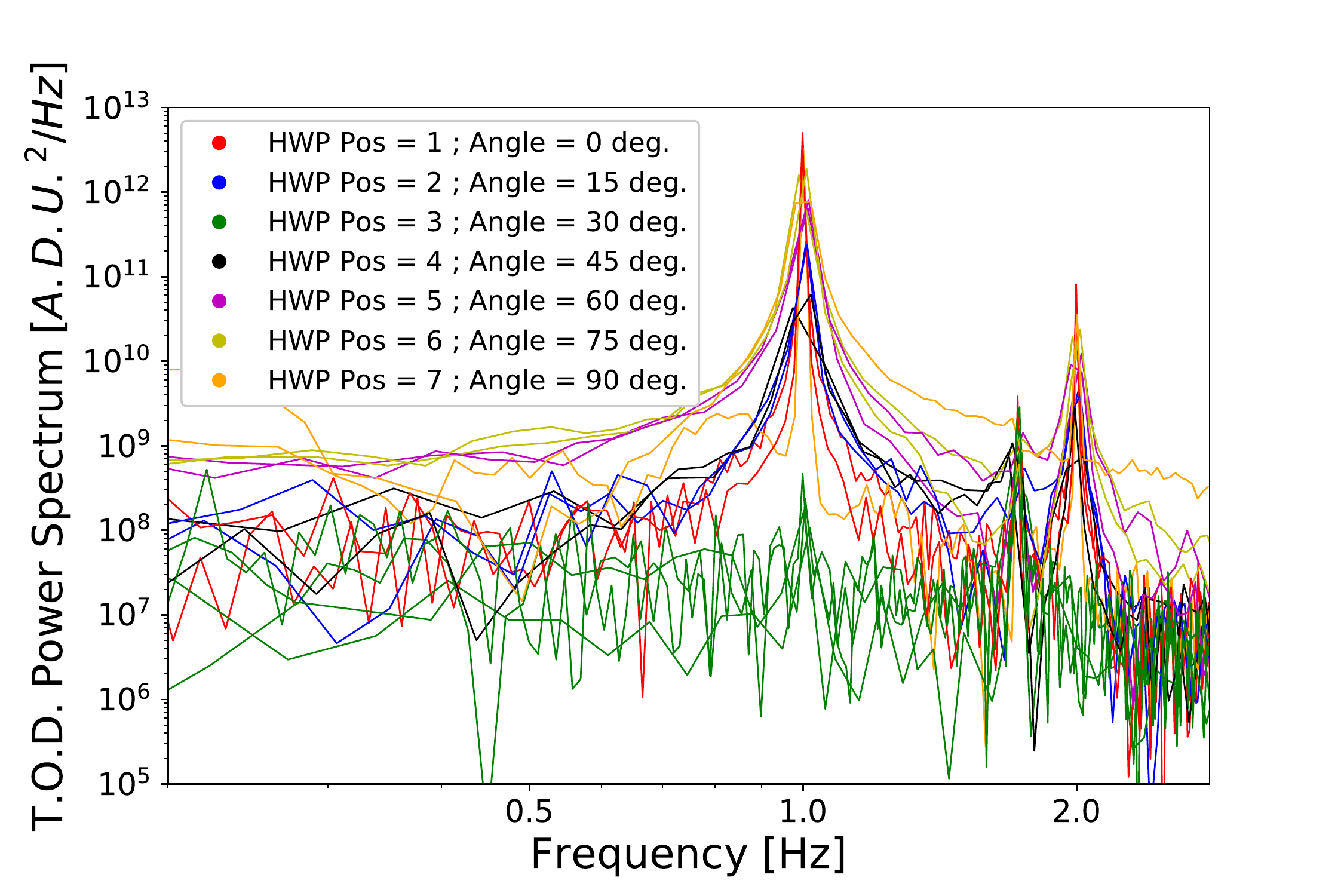}
\includegraphics[width=70mm,height=60mm]{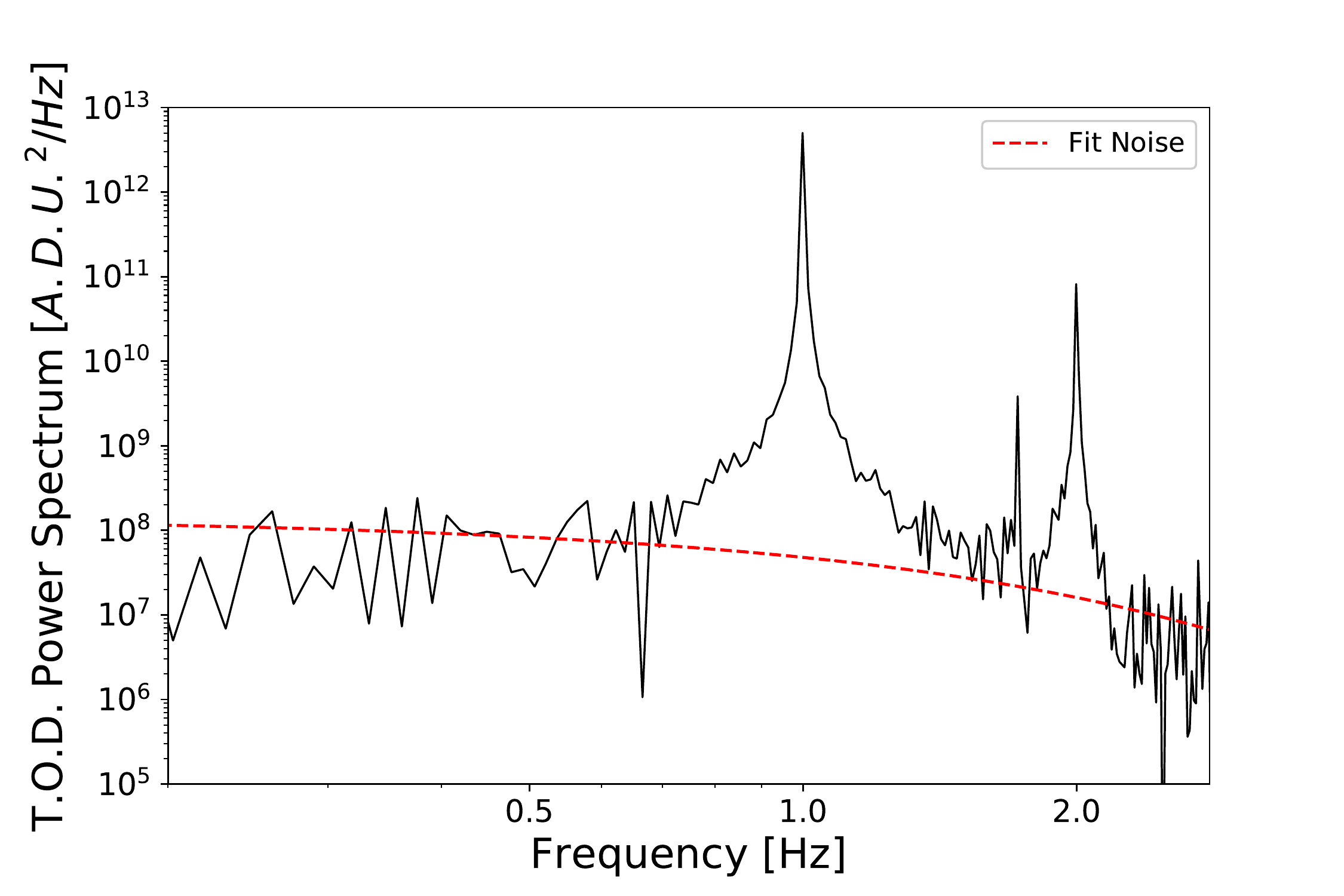}
\caption{\textit{Left}: power spectra of the TOD in Arbitrary Digital Units at each HWP position, reported with different colors. The spectra at the $3^{rd}$ HWP position correspond to data nearby the minimum of the polarization modulation curve and are therefore covered by the noise. \textit{Right}: power spectrum of TOD acquired at the $1^{st}$ HWP position, the red dashed line shows the exponential fit on the data deprived of the main band signal, used for the noise estimation.
The calibration source signal emits at \SI{1}{\hertz} as expected and at about \SI{1.7}{\hertz} the pulse tube signal can be seen in the spectra as a vibration effect on the detectors and a lower amplitude calibration source harmonics at \SI{2}{\hertz} can be seen as well.}
\label{fig:powspe}
\end{figure}

\begin{figure}[ht]
\centering
\includegraphics[scale=0.5]{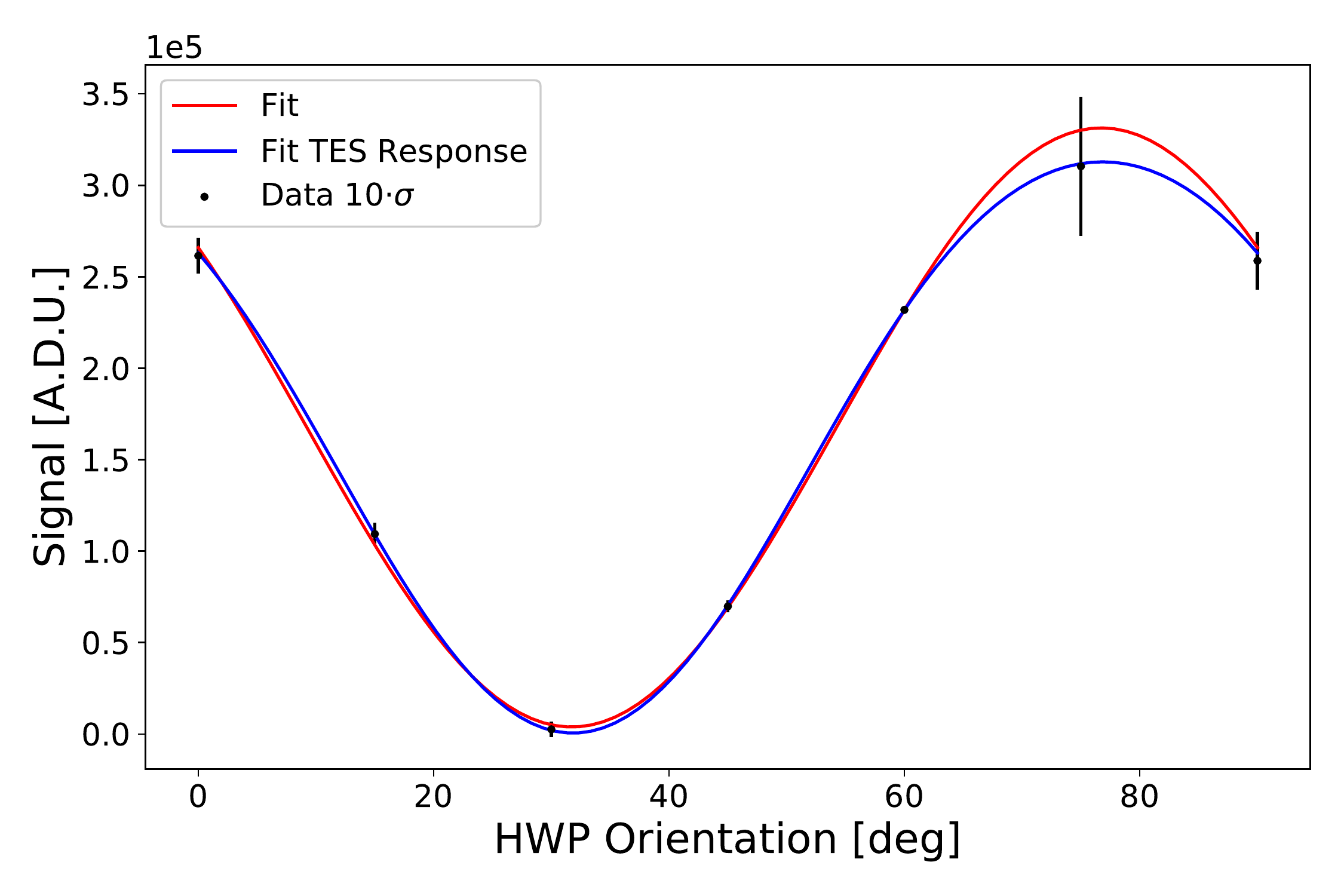}
\caption{Polarization modulation curve acquired at the nominal HWP positions $[\SI{0}{\degree}, \SI{15}{\degree},\SI{30}{\degree},\SI{45}{\degree},\SI{60}{\degree},\SI{75}{\degree},\SI{90}{\degree}]$, reported in Arbitrary Digital Units. Data points are reported as mean values and errors ($10\sigma$ error bars) on the mean of the data processed by the power spectra integration over the main band. Two models are used to fit the data: a model as in Eq.~\ref{eq:fit} (red line) and a similar model where the response of the detector is included as a function which deviates from linearity for high input power, as in Eq.~\ref{eq:fit1}  (blue line).}
\label{fig:polmod}
\end{figure}

The QUBIC polarimeter has been tested during the calibrations, with the focal plane cooled down to $\SI{348}{\milli\kelvin}$ and with the HWP rotator at a temperature of $\SI{8.5}{\kelvin}$. The experimental setup for the polarization calibration consists of a \SI{150}{\giga\hertz} calibration source placed \SI{11}{\meter} away from the cryostat, which was tilted \cite{2020AA.QUBIC.PAPER3}. The source points at a flat mirror, $\SI{24.4}{\degree}$ tilted, which redirects the radiation orthogonally to the UHMW\cite{2018InPhT..90...59D} window of the cryostat. \\
The calibration source emits fully polarized radiation which is modulated at $\SI{1}{\hertz}$. Here we report the data acquired by a representative detector ($\textit{TES}~ \#95$). Data are acquired as time ordered data (TOD) at the seven nominal rotator positions and are processed via Fourier transform, providing an estimate of the signal by integrating the power spectra in the main band $[0.7, 1.3]\SI{}{\hertz}$. The power spectra of each TOD are reported in the \textit{left panel} of \figurename~\ref{fig:powspe} with different colors for each HWP position. The noise is estimated by performing an exponential fit of the power spectra, deprived of the main band signal. This provides the noise power spectrum, reported as the red dashed line in the \textit{right panel} of \figurename~\ref{fig:powspe}, and the noise contribution to the signal is in the end estimated by integrating the noise power spectrum in the same band of the main signal and subtracted.
The polarization modulation curve is reported in \figurename~\ref{fig:polmod}, where the mean values and the errors on the mean of the processed data are reported as a function of the nominal HWP positions. 
In the same figure the fit of the data is reported in red. The fit has been performed with a function: 
\begin{equation}
I = \frac{1}{2}\left[T+Q\cos(4\theta + \phi)\right]   
\label{eq:fit}
\end{equation}
where an almost pure linear polarized light is assumed as input (\textit{VDI electronics} Gunn oscillator 130-\SI{170}{\giga\hertz}) but with still unknown polarization angle, driven by the setup alignment. An offset $\phi$ on the HWP orientation is included in the fit model such that, if expanded, provides the same relation as in Eq.~\ref{eq:ideal_polarimeter}.
The retrieved fit parameters show an almost pure linearly polarized input radiation.
The detector response needs to be included for the data nearby the maximum of the polarization modulation curve, which clearly deviate from the expected sine wave function \cite{2020AA.QUBIC.PAPER3}. In order to consider the deviation from the linear response of the detector for high values of the input power, a suitable fit model is used with a function :
\begin{equation}
I'=k\cdot tanh(I/k)
\label{eq:fit1}
\end{equation}

which grows linearly with the signal in the limit $I\ll k$ and tends to $k$ for $I\gg k$, where $I$ is the modulated intensity of Eq.~\ref{eq:fit}. This provides the final estimate of the input Stokes parameters from the polarization modulation curve provided by the QUBIC Stokes polarimeter. The polarization modulation, together with the fit models, is reported in \figurename~\ref{fig:polmod}. Measurements of the TES linearity have only been performed when the Stokes polarimeter was not yet assembled within the QUBIC cryostat, and therefore cannot be included in the data analysis process. Linearity measurements will be performed again with the inclusion of the half-wave plate polarimeter.
In \tablename~\ref{tab:6.1} the best fit parameters of the two models used to fit the data are reported, showing an almost pure input linear polarization retrieved from the measured polarization modulation curve. The obtained results at 150GHz are compatible with the very low cross-polarization estimated in \cite{2020AA.QUBIC.PAPER3}.

\begin{table}
\centering
\caption{Best fit parameters of the measured polarization modulation curve, retrieved assuming a model as in Eq.~\ref{eq:fit}, and having also included in the fit model a suitable function which considers the detector response for high input power, as in Eq.~\ref{eq:fit1}. The retrieved Stokes parameters, in Arbitrary Digital Units (\textit{A.D.U.}), show an almost pure input linear polarization. These results are compatible with a low level of cross-polarization as estimated in \cite{2020AA.QUBIC.PAPER3} in the same frequency band.}
\begin{tabular}[b]{cccc}
	\hline\hline
	Model & $T$ [A.D.U.] & $Q$ [A.D.U.]& $\phi$ [rad]\\
	\hline
	As in Eq.~\ref{eq:fit} & $(3.352\pm 0.035)\cdot 10^{5}$ & $(3.272\pm 0.042)\cdot 10^{5}$ & $0.927 \pm 0.013 $\\
	As in Eq.~\ref{eq:fit1} & $(3.620\pm 0.008)\cdot 10^{5}$ & $(3.624\pm 0.010)\cdot 10^{5}$ & $0.916 \pm 0.001$\\
	\hline

\end{tabular}
	\label{tab:6.1}
\end{table}

\section{\label{subse:impro}Full Instrument improvements}
Some important improvements are planned when the instrument will be upgraded from the Technological Demonstrator to the Full instrument. 
\begin{itemize}
    \item The stretch pulley will be actively tuned thanks to a small cryogenic actuator. This system is crucial to minimize the friction during the rotation. 
    \item The hourglass bearings will be replaced with fully ceramic cryogenic bearings.
    \item Hardware limit switches, suitable at cryogenic temperature, are under testing. Currently only a software limit switch has been implemented. 
\end{itemize}

\section{\label{sec:concl}Conclusion}

We developed a Stokes polarimeter able to operate at cryogenic temperature for the QUBIC experiment for the quest of CMB B-modes. The polarimeter, composed of a step-by-step rotating meta-material-HWP, has been tested several times at room temperature, in liquid Nitrogen, and within the QUBIC cryostat. A dedicated position readout system has been developed and tested, allowing to move and control the HWP with a precision of $\SI{0.1}{\degree}$. In this paper we show how the device allows to carry out measurements by limiting the heating of the HWP to values $\leq\SI{10}{\milli\kelvin}$, satisfying the low temperature requirement for this optical element.
We also report here \SI{150}{\giga\hertz} polarization measurements, proving the capability of retrieving the input Stokes vector of a purely polarized input light with extremely low cross-polarization, when the non-linearity of the detector is included. The measured optical model of the QUBIC Stokes polarimeter, in the form of Mueller matrices, is reported here at \SI{150}{\giga\hertz} and will be included in a future work to provide the performance of the QUBIC polarimeter at all the frequencies available for the QUBIC-TD, along with new measures to be taken.

\acknowledgments
QUBIC is funded by the following agencies. France: ANR (Agence Nationale de la Recherche) 2012 and 2014, DIM-ACAV (Domaine d’Interet Majeur-Astronomie et Conditions d’Apparition de la Vie), CNRS/IN2P3 (Centre national de la recherche scientifique/Institut national de physique nucle´aire et de physique des particules), CNRS/INSU (Centre national de la recherche scientifique/Institut national 8 Battistelli et al de sciences de l’univers). Italy: CNR/PNRA (Consiglio Nazionale delle Ricerche/Programma Nazionale Ricerche in Antartide) until 2016, INFN (Istituto Nazionale di Fisica Nucleare) since 2017. Argentina: MINCyT (Ministerio de Ciencia, Tecnología e Innovación), CNEA (Comisión Nacional de Energía Atómica), CONICET (Consejo Nacional de Investigaciones Científicas y Técnicas).

D. Burke and J.D. Murphy acknowledge funding from the Irish Research Council un- der the Government of Ireland Postgraduate Scholarship Scheme. D. Gayer and S. Scully acknowledge funding from the National University of Ireland, Maynooth. D. Bennett ac- knowledges funding from Science Foundation Ireland.

\afterpage{\clearpage}
\bibliographystyle{ieeetr}
\bibliography{qubic} 

\end{document}